\documentclass[12pt,aps,nofootinbib,floatfix]{revtex4}

\usepackage{amsmath}
\usepackage{amsfonts}
\usepackage{amssymb}
\usepackage{color}

\def\bc{\begin{center}}
\def\nno{\nonumber}
\def\ec{\end{center}}
\def\be{\begin{eqnarray}}
\def\ee{\end{eqnarray}}

\newcommand{\omits}[1]{}

\definecolor{dyellow}{rgb}{1.,0.8,.0}
\definecolor{myblue}{rgb}{.1,.1,.7}
\definecolor{dcyan}{rgb}{.0,.6,.6}
\definecolor{dmagenta}{rgb}{0.6,0.0,0.6}
\definecolor{brown}{rgb}{0.6,0.2,0.}
\definecolor{darkblue}{rgb}{.0,.0,0.5}
\definecolor{darkred}{rgb}{0.75,0.0,0.0}
\definecolor{orange}{rgb}{1.,.6,.0}
\definecolor{dorange}{rgb}{0.8,.4,.0}
\definecolor{darkgreen}{rgb}{0.0,0.6,0.0}
\definecolor{purple}{rgb}{.4,.0,.4}
\definecolor{lightgrey}{rgb}{0.7, 0.7, 0.7}
\definecolor{grey}{rgb}{0.4, 0.4, 0.4}

\def\Si{\Sigma}

\def\al{\alpha}
\def\ga{\gamma}

\def\la{\lambda}

\def\Ga{\Gamma}

\newcommand\btd{\raise 2pt
\hbox{$\hat\bigtriangledown$}\hskip 1.5pt}
\newcommand\bt{\raise 2pt
\hbox{$\bigtriangledown$}\hskip 1.5pt}

\def\NPB{{Nucl. Phys.}~{\bf B}}
\def\PRD{{Phys. Rev.}~{\bf D}}
\def\PRL{{Phys. Rev. Lett. }}

\def\PLB{{Phys. Lett.}~{\bf B}}

\begin{document}

\title{Note on Chern-Simons Term Correction to Holographic Entanglement Entropy}

\author{{Jia-Rui Sun}}
\email{sun@ihep.ac.cn}
\affiliation{
   Institute of High Energy Physics, Chinese Academy of Sciences,
   P.O. Box 918-4, Beijing 100049, China}
\affiliation{
   Graduate School of Chinese Academy of Sciences,
   Beijing 100049, China}


\vskip 10cm

\begin{abstract}

From the perspective of AdS/CFT correspondence, we study the
gravitational Chern-Simons term correction to the holographic
entanglement entropy of CFT on the conformal boundary of
asymptotically AdS$_3$ spacetime using the off-shell Euclidean path
integral method. We show that, like the BTZ black hole entropy, the
holographic entanglement entropy is indeed modified  due to the
gravitational Chern-Simons term although the bulk geometry does not
change. 

\end{abstract}

\maketitle

\newpage

\tableofcontents

\section{Introduction}

The entanglement entropy (often defined as the von Neumann entropy),
which is used to describe the correlations between two (or more)
subsystems of a quantum field system, has received a lot of
attention over the past few years. It plays an important role in
studying the critical phenomena or phase transitions in condensed
matter physics and quantum field theory (for a recent review, see,
e.g., \cite{cardy}). Since it also describes the lack of information
for an observer in a subsystem $A$ who is inaccessible to other
subsystems, the concept of entanglement entropy is also applied to
black hole physics, where the black hole entropy is viewed as an
entanglement entropy resulting from the entanglement of the vacuum
states of certain quantum fields inside and outside the event
horizon of black hole~\cite{bombelli, callan, holzhey, jacobson}.

From the prospective of holography \cite{'t Hooft, susskind,
Polyakov, Witten, Maldacena}, the entanglement entropy acquires new
understanding. In Ref.~\cite{rt}, Ryu and Takayanagi proposed the
concept of holographic entanglement entropy (HEE): the entanglement
entropy of boundary CFT can be calculated by considering bulk
geometry of AdS space. They considered CFT on an $d$-dimensional
flat spacetime which serves as the (conformal) boundary of an
$(d+1)$-dimensional bulk anti-de Sitter spacetime (AdS$_{d+1}$) and
proposed  that the entanglement entropy $S_A$ of an $d$-dimensional
CFT system in a region $A$ in the flat spacetime is determined by an
$(d-1)$-dimensional spacelike, static, minimal hypersurface
$\gamma_A$ in the bulk with $\partial A$ as its boundary via
\begin{equation}\label{hee}
S_A=\frac{\mathcal{A}}{4G_{d+1}} ,
\end{equation}
where the conformal field in the region $A$ is a subsystem of the
whole system in $A+B$ in the flat spacetime, $\mathcal{A}$ is the
area of $\gamma_A$, and $G_{d+1}$ is the $(d+1)$-dimensional Newton
gravitational constant. Note that Eq.(\ref{hee}) has the same form
as the black hole area entropy formula. This is the reason why they
call $S_A$ the holographic entanglement entropy. In Ref.~\cite{rt},
they checked the validity of this relation by a few examples. For
the case of AdS$_3$/CFT$_2$ correspondence, the entanglement entropy
calculated from the well-known 2-dimensional CFT matches
Eq.(\ref{hee}) very well. When it is applied to higher dimensions,
e.g., AdS$_5$/CFT$_4$, however, the entropies calculated from both
sides are not exactly the same. It is argued that the occurrence of
this discrepancy is due to the dual description of the classical
gravity in bulk AdS is a strongly coupled supersymmetric Yang-Mills
theory, but not a free one, while they only considered the free
field approximation in the side of field theory. This is similar to
the case of the well-known difference by a factor $3/4$ between the
entropy of black D$3$-branes and the entropy of free supersymmetric
Yang-Mills theory.

Shortly, the proposal has been extended to many cases such as
calculating the entropy of black holes~\cite{emparan}, de Sitter
brane world~\cite{dsbrane} and studying the
confinement/deconfinement transition of gauge fields~\cite{kkm}, and
others, see for example Ref.~\cite{hee}. In the mean time, the
proposal has been studied in more detail in~\cite{add,covhee} and
Fursaev~\cite{proofhee} has given a proof for the proposal by
applying the Euclidean path integral approach~\cite{path} to Riemann
surfaces with conical singularities~\cite{offshell} in the AdS/CFT
correspondence.

Note that the proposal relates the bulk geometry to the HEE of
boundary CFT and in the AdS/CFT corresponence, the higher order
curvature terms in the bulk are dual to the large $N$ expansion
corrections in the dual CFT, it is natural to consider higher order
corrections to the bulk gravitational action and to see how the HEE
of boundary CFT gets modified. Typical higher order corrections to
the Hilbert-Einstein action are Gauss-Bonnet term~\cite{GB} and
Chern-Simons term~\cite{Chern} in the low energy effective action of
string theory. The Gauss-Bonnet term correction to HEE has been
briefly discussed in \cite{proofhee}. Now an interesting question is
to see how the Chern-Simons term correction modifies the HEE of dual
CFT. It is interesting because in general, the presence of the
gravitational Chern-Simons term modifies the bulk equation of motion
by the Cotton tensor. However, for the global AdS$_3$ spacetime and
the BTZ black hole, the Cotton tensor vanishes identically, i.e.,
the bulk geometry does not change and a direct use of Eq.(\ref{hee})
leads no modification. It does not mean that the Chern-Simons term
correction is trivial in this case, because it violates the
diffeomorphism invariance of the bulk gravitational action. In the
AdS/CFT correspondence, the violation of diffeomorphism invariance
in asymptotically AdS$_3$ (AAdS$_3$) spacetime causes gravitational
anomaly (covariant anomaly) in the dual 2-dimensional CFT on the
conformal boundary and changes the central charges of the CFT, this
corresponds to a modification to the BTZ black hole entropy by a
term proportional to its inner horizon \cite{kraus,solo,park1}. The
gravitational Chern-Simons term correction to black hole entropy has
also been studied in \cite{sen,tachi} using the Wald's entropy
formula~\cite{wald}. Although in this case the variation of the
Chern-Simons term doesn't modify the bulk geometry, it does modify
the black hole entropy. The relation between trace anomalies and HEE
has been studied in \cite{theisen}.

The purpose of this paper is to study how the gravitational
Chern-Simons term modifies the HEE in AAdS$_3$ spacetime. The
standard way to calculate its entanglement entropy is the replica
trick \cite{cardy} which computes the partition function of the
quantum fields in Euclidean path integral formulation on an
$n$-sheeted Riemann surface, in the limit of $n \rightarrow 1$. The
conical singularity is introduced as a result of a conformal
coordinate transformation on the $n$-sheeted Riemann surface, which
produces a $\delta$ function in the Riemann tensor. Then using the
off-shell Euclidean path integral method \cite{proofhee}, we find
that to the leading order the gravitational Chern-Simons term indeed
varies the HEE (when the bulk is BTZ black hole) as expected. We
point out here that this is a approximately calculation since we do
not actually solve the constraint condition for $\ga_A$.

The paper is organized as follows. In Sec.~II, we give a brief
introduction of the replica trick and the off-shell method
introduced in~\cite{cardy,proofhee}. In Sec.~III we calculate the
gravitational Chern-Simons term correction to the HEE in the cases
of zero temperature and finite temperature, respectively. Then in
Sec.~IV we make some conclusions and discussions.

\section{The replica trick and the off-shell method}
\subsection{Replica trick}

Let's consider an $d$-dimensional CFT $\phi(x)$ defined on a flat
spactime $\mathbb{R}^{1,d-1}$. Divide the CFT system into two
subsystems $A$ and $B$ with boundary $\partial_A$. The density
matrix $\rho$ of the whole system $A+B$ in a thermal state at
inverse temperature $\beta$ can be expressed in the Euclidean path
integral as~\cite{cardy}
\be\label{rho} \rho_{\phi''\phi'}=\frac{1}{Z[\beta]}\int
\mathcal{D}\phi(x)\prod_x
\delta(\phi(\vec{x},0)-\phi'(\vec{x},0))\prod_x
\delta(\phi(\vec{x},\beta)-\phi''(\vec{x},0))e^{-I_E}, \ee
where $I_E$ is the Euclidean action of the whole system and the
unitarity requires $tr\rho=1$. Let the subsystem $A$ consists of
points $x$ in some region. The reduced density matrix $\rho_A(={\rm
tr}_B\rho)$ for the subsystem $A$ is obtained by sewing together the
points which only belong to $B$. Then the entanglement entropy of
$A$ is calculated from the von Neumann entropy
\be S_A=- {\rm tr}_A \rho_A \ln \rho_A, \ee
To compute $S_A$, define a new quantity ${\rm tr }\rho_A^n\equiv
Z_n(A)/Z^n$, $Z_n(A)$ is the partition function over an $n$-sheet
structure by making $n$ copies of the fields and sewing them
together through $\phi'(x)_k=\phi''(x)_{k+1}$ and
$\phi'(x)_1=\phi''(x)_n$ ($1\leq k\leq n$) \footnote{As has been
mentioned in \cite{theisen}, there is another sewing condition which
needs $\phi(x)$ to be continuous at boundary $\partial_A$.}. Then
$S_A$ is given by
\be \label{sa}
S_A=-\lim_{n\rightarrow 1}\frac{\partial}{\partial
n}\frac{Z_n(A)}{Z^n}.
 \ee
Note that in the limit of $n \rightarrow 1$, Eq.~(\ref{sa}) can be
expressed as the Tsallis entropy~\cite{tsallis}
\be S_A=\lim_{n\rightarrow 1}\frac{ {\rm tr} \rho_A^n-1}{1-n}. \ee

\subsection{Off-shell method}

Following Subsection A, the $d$-dimensional CFT is defined on
\be \label{nflat} ds^2=d\tau^2+dx^2+h_{ij}dy^idy^j,
\ee
where $\tau$ is the Euclidean time, $h_{ij}$ is the spatial
component of the metric and $i,j$ $\in$ \{2,3,$\cdots$, $d-1$\}.
Recall that in 2-dimensional space, the conformal transformations
can be realized as the holomorphic coordinate transformations in
complex plane \cite{ginsparg}. So we can define
\be w=\tau+ix \quad {\rm and}\quad \bar{w}=\tau-ix. \ee
The $n$-sheet structure mentioned above is realized by a singular
coordinate transformation~\cite{theisen, micha}
\be\label{w} w\rightarrow z=w^{\frac{1}{n}}, \ee
it is singular at $z=0$ (when $n\neq 1$), which is the boundary
between A and B at $\tau=0$. Then Eq.(\ref{nflat}) becomes
\be\label{singularm}
ds^2&=&n^2z^{n-1}\bar{z}^{n-1}dzd\bar{z}+h_{ij}dy^idy^j.
\ee
Let $n=1+\epsilon$, $\epsilon$ is an infinitesimal real parameter.
To the linear order of $\epsilon$, Eq.(\ref{singularm}) is
\be\label{epsilon} ds^2&=&(1+2\epsilon+\epsilon
\ln(z\bar{z}))dzd\bar{z}+h_{ij}dy^idy^j+\mathcal{O}(\epsilon^2)\nno\\
&\equiv&2h_{z\bar{z}}dzd\bar{z}+h_{ij}dy^idy^j+\mathcal{O}(\epsilon^2).
\ee
A direct calculation shows that the Riemann tensor contains singular
components due to the $\ln(z\bar{z})$ term in Eq.(\ref{epsilon})
\be \Ga^{\bar{z}}_{\bar{z}\bar{z}}=h^{z{\bar{z}}}\frac{\partial
h_{z{\bar{z}}}}{\partial \bar{z}} \quad {\rm and}\quad
\Ga^z_{zz}=h^{z{\bar{z}}}\frac{\partial h_{z{\bar{z}}}}{\partial
z},\\ \nno \ee
\be\label{riemann} R^z_{zz\bar{z}}&=&\frac{\partial
\Ga^z_{z\bar{z}}}{\partial z}-\frac{\partial
\Ga^z_{zz}}{\partial\bar{z}}+\Ga^z_{zz}\Ga^z_{z\bar{z}}-\Ga^z_{z\bar{z}}\Ga^z_{zz}=-\frac{\partial
\Ga^z_{zz}}{\partial\bar{z}}, \nno\\
R^{\bar{z}}_{\bar{z}\bar{z}z}&=&\frac{\partial
\Ga^{\bar{z}}_{z\bar{z}}}{\partial \bar{z}}-\frac{\partial
\Ga^{\bar{z}}_{\bar{z}\bar{z}}}{\partial
z}+\Ga^{\bar{z}}_{\bar{z}\bar{z}}\Ga^{\bar{z}}_{z\bar{z}}-\Ga^{\bar{z}}_{z\bar{z}}\Ga^{\bar{z}}_{\bar{z}\bar{z}}
=-\frac{\partial \Ga^{\bar{z}}_{\bar{z}\bar{z}}}{\partial z}.\ee
The singular part of Riemann tensor is
\be \label{rs} ^sR^z_{zz\bar{z}}&=&
^sR^{\bar{z}}_{\bar{z}\bar{z}z}=-2\pi\epsilon\delta^{(2)}(z,\bar{z}),\ee
where the following relations have been used
\be\label{sr}
R^z_{zz\bar{z}}&=&R_{z\bar{z}}=-h^{z\bar{z}}\frac{\partial^2
h_{z\bar{z}}}{\partial z\partial \bar{z}}-\frac{\partial
h^{z\bar{z}}}{\partial \bar{z}}\frac{\partial h_{z\bar{z}}}{\partial
z}  \nno\\
&=&-2(1-2\epsilon-2\epsilon\ln(z\bar{z}))\frac{\epsilon}{2}2\pi\delta^{(2)}(z,\bar{z})
-\frac{\partial h^{z\bar{z}}}{\partial \bar{z}}\frac{\partial
h_{z\bar{z}}}{\partial z} \nno\\ {\rm and}\quad \frac{\partial^2
\ln(z\bar{z})}{\partial z\partial
\bar{z}}&=&2\pi\delta^{(2)}(z,\bar{z}).\ee
Then the singular part of Ricci scalar is
\be\label{riccis} ^sR=2h^{z\bar{z}}R_{z\bar{z}}=-4\pi\epsilon
h^{z\bar{z}}\delta^{(2)}(z,\bar{z}). \ee

Now the key point is that in the AdS/CFT correspondence
\cite{Maldacena,Polyakov,Witten}, the partition functions of the
bulk (super)gravity $Z_{gr}(h)$ and boundary CFT $Z_{CFT}(h)$ can be
related through
\be\label{zz} Z_{gr}(h)=Z_{CFT}(h),\ee
and $Z_{gr}(h)$ can be calculated from its Euclidean path integral
as $Z_{gr}(h)=\int\mathcal{D}[g]\exp(-I_E(g))$, where $g$ is the
bulk metric with induced metric $h$ on the boundary. For
$3$-dimensional Einstein gravity with a negative cosmological
constant, the Einstein-Hilbert action with Gibbons-Hawking surface
term is
\be \label{EH} I_{gr}&=&\frac{1}{16\pi
G}\int_{\mathcal{M}}\sqrt{-g}d^3x(R+\frac{2}{l_a^2})+\frac{1}{8\pi
G}\int_{\mathcal{\partial M}}\sqrt{|h|}d^2x K, \ee
where $l_a$ is the radius of AdS$_3$ spacetime and $K$ is the
extrinsic curvature on the boundary $\partial {\mathcal M}$. The
dual field theory on $\partial {\mathcal M}$ is a 2-dimensional CFT
on a circle consisting of subsystems $A$ and $B$. Thanks to the
$n$-sheeted structure on the boundary, it introduces a discrete
parameter $\beta=2\pi n$ in the partition function which reads
\cite{proofhee}
\be Z_{CFT}(h,T)=\lim_{\beta\rightarrow 2\pi}Z_{CFT}(\beta,h,T), \ee
where $T$ is the temperature of the CFT system. Analogous to the
treatment in canonical ensemble, the entanglement entropy of A is
given by \cite{fursaev}
\be\label{sa1} S_A=\lim_{\beta\rightarrow
2\pi}(1-\beta\frac{\partial}{\partial\beta})\ln
Z_{CFT}(\beta,h,T).\ee
With Eq.(\ref{zz}), $S_A$ can be calculated by
\be\label{sa2} S_A&=&\lim_{\beta\rightarrow
2\pi}(1-\beta\frac{\partial}{\partial\beta})\ln Z_{CFT}(h) \nno\\
&=&\lim_{\beta\rightarrow
2\pi}(1-\beta\frac{\partial}{\partial\beta})(-I^{(0)}_E(g_0)-I^{(2)}_E(g_0)+\cdots),
\ee
where $I^{(0)}_E(g_0)$ indicates the (dominant) contribution from
the solution $g_0$ (of bulk dynamical equation), and
$I^{(1)}_E(g_0)=0$, i.e., bulk dynamical equation has been used. In
the saddle point approximation (zero loop approximation),
Eq.(\ref{sa2}) can be well approximately computed as
\be\label{sa3} S_A&=&\lim_{\beta\rightarrow
2\pi}(1-\beta\frac{\partial}{\partial\beta})(-I^{(0)}_E(g_0)). \ee
Dividing the the gravitational action (\ref{EH}) into the regular
part and singular part and using Eq.(\ref{riccis}) we obtain
\be \label{EH1} iI_{gr}=-I_E&=&\frac{i}{16\pi
G}\int_{\mathcal{M}}\sqrt{-g}d^3x(
^rR+\frac{2}{l_a^2})+\frac{i}{8\pi G}\int_{\mathcal{\partial
M}}\sqrt{|h|}d^2x K  \nno \\  &&+ \frac{1}{16\pi
G}\int_{\mathcal{M}}h_{z\bar{z}}\sqrt{\ga}dz d\bar{z}dx
 ^sR  \nno \\
&=& \frac{i}{16\pi G}\int_{\mathcal{M}}\sqrt{-g}d^3x(
^rR+\frac{2}{l_a^2})+\frac{i}{8\pi
G}\int_{\mathcal{\partial M}}\sqrt{|h|}d^2x K \nno \\
&&-\frac{\epsilon}{4G}\int_\Si\sqrt{\ga}dx, \ee
where $\ga$ is the induced metric of the spatial section $\Si$ with
codimension-2. The term $\int_\Si\sqrt{\ga}dx$ is the
Dirac-Nambu-Goto like action which denotes the area $\mathcal{A}$ of
the codimension-2 surface. Then we obtain
\be\label{sa4}
S_A=\frac{1}{4G}\int_\Si\sqrt{\ga}dx=\frac{\mathcal{A}}{4G}.\ee
The constraint equation $\delta \mathcal{A}=0$ requires
$\mathcal{A}$ be the bulk minimal surface with the boundary
$\partial_A$. Eq.(\ref{sa4}) is just of the form of Eq.(\ref{hee})
in AdS$_3 $ spacetime. The above discussion can be generalized into
higher dimensional spacetimes.

\section{Gravitational Chern-Simons term corrections to HEE}

The action of 3-dimensional Einstein gravity with a negative
cosmological constant in the presence of a gravitational
Chern-Simons term is
\be
I&=&I_{EH}+I_{CS}\nno \\
&=&\frac{1}{16\pi
G}\int_{\mathcal{M}}\sqrt{-g}d^3x(R+\frac{2}{l_a^2})+\frac{1}{8\pi
G}\int_{\mathcal{\partial M}}\sqrt{|h|}d^2x K
\nno\\&&+\frac{\al}{32\pi G}\int_{\mathcal{M}}\sqrt{-g}d^3x
(\Ga^{\al}_{\beta \mu} \frac{\partial\Ga^{\beta}_{\al \la}}{\partial
x^{\nu}}+\frac{2}{3} \Ga^{\al}_{\beta \mu}\Ga^{\beta}_{\ga
\nu}\Ga^{\ga}_{\al\la})\epsilon^{\mu\nu\la}, \ee
where $\al$ is the coupling constant, $h$ is the induced metric on
the boundary, and $K$ is the extrinsic curvature of the boundary.
Gravity described by this action is called the topologically massive
gravity (TMG) \cite{Chern} and it recently received much attention
\cite{t,t1,t2,gj,gjj,park,t3,t4,t5,t6,t7}. Variation of $I$ yields
equation of motion
\be
R_{\al\beta}-\frac{1}{2}g_{\al\beta}R-\frac{1}{l_a^2}g_{\al\beta}+\al
C_{\al\beta}=0,\ee
where
\be C_{\al\beta}=\epsilon_{\al}
^{~\mu\nu}\bigtriangledown_{\mu}(R_{\nu\beta}-\frac{1}{4}g_{\nu\beta}R)
\ee
is called the Cotton tensor.

TMG permits the pure AdS$_3$ spacetime and BTZ black hole as its
solutions and the Cotton tensor vanishes identically in these cases.
Here we only focus on these cases.

For the AdS$_3$ spacetime and BTZ black hole, a direct use of
eq.(\ref{hee}) reflects no correction to HEE in the presence of the
gravitational Chern-Simons term. Since the bulk geometries do not
change in those cases. However, it does not mean that the
Chern-Simons term correction is trivial in these examples. Because
the presence of gravitational Chern-Simons term violates the
diffeomorphism invariance of the gravitational action. In the
AdS/CFT correspondence, the gravitational anomaly (covariant anomly)
occurs in the dual 2-dimensional CFT and modifies its central
charges as \cite{kraus}
\be\label{c}c_L=c(1+\frac{\al}{l_a})\quad {\rm and}\quad
c_R=c(1-\frac{\al}{l_a}),\ee
where $c_L$ and $c_R$ are the central charges of the left-moving and
right-moving sectors of the CFT, respectively, and  $c=3l_a/(2G)$.
The microscopic entropy of the CFT is computed from Cardy's formula
\be\label{sbtz}S_\mathrm{CFT}&=&2\pi\left(\sqrt{\frac{h_L
c_L}{6}}+\sqrt{\frac{h_R c_R}{6}}\right)=\frac{\pi^2}{3}(c_L T_L+c_R
T_R),\ee
where $h_L=\frac{Ml_a-J}{2}$ and $h_R=\frac{Ml_a+J}{2}$ are the
conformal weights of the boundary stress tensor,
$T_L=\frac{r_+-r_-}{2\pi l_a}$ and $T_R=\frac{r_++r_-}{2\pi l_a}$
are the left-hand and right-hand temperatures of the dual
2-dimensional CFT, $r_+$ and $r_-$ are outer and inner horizons of
the BTZ black hole, and
\be
M=m-\frac{\al}{l_a^2}j=\frac{r_+^2+r_-^2}{8Gl_a^2}-\frac{\al}{l_a^2}\frac{r_+r_-}{4Gl_a}\quad
{\rm and}\quad J=j-\al m=\frac{r_+r_-}{4Gl_a}-\al
\frac{r_+^2+r_-^2}{8Gl_a^2}\ee
are the mass and angular momentum of the BTZ black hole in the
presence of the gravitational Chern-Simons term. As a result, it can
be seen that the gravitational anomaly is reflected by a
modification to the BTZ black hole entropy by a term proportional to
its inner horizon \cite{solo}
\be S_\mathrm{BTZ}&=&S_\mathrm{CFT}=\frac{\pi r_+}{2G}-\frac{\pi\al
r_- }{2Gl_a}=\frac{\pi c}{3l_a}(r_+-\frac{\al r_-}{l_a}).\ee
The same result is obtained using other methods \cite{sen,tachi}.

Note that in the AdS/CFT correspondence, the BTZ black hole entropy
is identical to the entropy of its dual 2-dimensional CFT on the
conformal boundary at spatial infinity. Thus it is interesting to
consider how
 the gravitational Chern-Simons term modifies the HEE of dual CFT.
 To see this,
recall that the solutions of Einstein equation which are AAdS
spacetime can be expanded through the Fefferman-Graham (FG)
expansion in the Gaussian normal coordinates
\cite{fg,kraus,skenderis1,skenderis2}
\be\label{Gauss} ds^2=d\rho^2+l_a^2h_{ij}dx^idx^j,\ee
where
\be\label{fg}
h_{ij}=e^{2\rho/l_a}h^{(0)}_{ij}+h^{(2)}_{ij}+e^{-2\rho/l_a}h^{(4)}_{ij}+\cdots,
\ee
The conformal boundary of AAdS spacetime is located at
$\rho\rightarrow \infty$, and the induced metric on it is
$h^{(0)}_{ij}$. The stress energy tensor of the dual 2-dimensional
CFT is \cite{kraus,kraus1,kraus2}
\be \label{stress}T_{ij}=\frac{1}{8\pi
Gl_a}(h^{(2)}_{ij}-h_{(0)}^{kl}h^{(2)}_{kl}h^{(0)}_{ij}).\ee
It can be seen that although the CFT is defined on the boundary with
metric $h^{(0)}_{ij}$, its stress energy tensor relates to the bulk
metric up to $h^{(2)}_{ij}$. Besides, the contribution from the
gravitational Chern-Simons term to $T_{ij}$ also include
$h^{(2)}_{ij}$ terms as
\be\label{stress2}T_{ij}=\frac{1}{8\pi
Gl_a}(h^{(2)}_{ij}-h_{(0)}^{kl}h^{(2)}_{kl}h^{(0)}_{ij})-\frac{\al}{16\pi
Gl_a^2}(h^{(2)}_{ik}\epsilon_{lj}h^{kl}_{(0)}+h^{(2)}_{jk}\epsilon_{li}h^{kl}_{(0)}).
\ee
This indicates that to study the gravitational Chern-Simons term
correction to HEE, we need to take the contribution of
$h^{(2)}_{ij}$ into consideration. It is shown that like the HEE,
the correction can also be expressed as the geometric quantities
like the metric, Christoffel symbol and the intrinsic curvature.

As we know that the low energy effective theory of string theory
contains a series of higher order derivative terms corrections,
these higher order derivative terms are the loop or quantum
corrections to the theory. From the dictionary of the AdS/CFT
correspondence, the bulk higher derivative corrections should
correspond to the quantum excitations of the dual quantum fields on
the boundary. It is rather straightforward to apply the replica
trick and the off-shell Euclidean path integral method to study the
more general higher curvature corrections to the HEE, for example,
the HEE in the Lovelock gravity. In this case, the higher curvature
corrections to the HEE is expected to take the form of a polynomial
of the Riemann scalar curvatures of the bulk minimal surface.
Another interesting question is to study how to incorporate the bulk
gauge fields contribution to the HEE of the dual CFT. In the rest of
the paper, we will focus on the Chern-Simons term correction to the
HEE of the 2d CFT on the conformal boundary of the asymptotical
AdS$_3$ spacetime.

\subsection{Zero temperature CFT case}

 The zero temperature 2-dimensional CFT case corresponds to the bulk global
AdS$_3$ spacetime with metric
\be\label{gads}ds^2&=&d\rho^2+l_a^2(-\cosh^2\frac{\rho}{l_a}
dt^2+\sinh^2\frac{\rho}{l_a} d\phi^2) \nno\\
&=&d\rho^2+e^{2\rho/l_a}(-\frac{l_a^2}{4}dt^2+\frac{l_a^2}{4}d\phi^2)-\frac{l_a^2}{2}dt^2
-\frac{l_a^2}{2}d\phi^2+e^{-2\rho/l_a}(-\frac{l_a^2}{4}dt^2+\frac{l_a^2}{4}d\phi^2).\ee
Its dual 2-dimensional CFT is defined on the boundary at large
$\rho=\rho_0\gg l_a$ with conformal metric $h^{(0)}_{ij}$, $\rho_0$
is related to the UV cutoff $a$ of the CFT. To use the replica trick
in Sec.~II, we work in the Euclidean signature by taking
$t\rightarrow -i\tau$. Then define
\be w=\frac{l_a}{2}\tau+i\frac{l_a}{2}\phi=z^n=z^{1+\epsilon},\ee
As has been mentioned above that we need to take into account of the
$h^{(2)}_{ij}$ terms. So in the $\rho=\rho_0\gg l_a$ limit
Eq.(\ref{gads}) can be approximately described by
\be\label{fggads}
ds^2&\simeq&d\rho^2+e^{2\rho/l_a}(1+2\epsilon+\epsilon
\ln(z\bar{z}))dzd\bar{z} \nno\\&&+(1+2\epsilon+2\epsilon\ln z)dz^2
+(1+2\epsilon+2\epsilon\ln\bar{z})d\bar{z}^2+\mathcal{O}(\epsilon^2)
\nno\\
&=&d\rho^2+2e^{2\rho/l_a}h_{z\bar{z}}dzd\bar{z}+h_{zz}dz^2+h_{\bar{z}\bar{z}}d\bar{z}^2
+\mathcal{O}(\epsilon^2)\nno\\&\equiv&d\rho^2+2g_{z\bar{z}}dzd\bar{z}+h_{zz}dz^2+h_{\bar{z}\bar{z}}d\bar{z}^2
+\mathcal{O}(\epsilon^2).
\ee
Thus we obtain
\be\label{gagads}
g&=&h_{zz}h_{\bar{z}\bar{z}}-e^{4\rho/l_a}h_{z\bar{z}}h_{\bar{z}z}\simeq
-\frac{e^{4\rho/l_a}}{4}(1+4\epsilon+2\epsilon\ln(z\bar{z}))+\mathcal{O}(\epsilon^2),
 \nno\\ \Ga^{\bar{z}}_{\bar{z}\bar{z}}&=&4(1-4\epsilon-2\epsilon\ln(z\bar{z}))h_{z{\bar{z}}}
\frac{\partial h_{z{\bar{z}}}}{\partial
\bar{z}}+\frac{1}{2}h^{\bar{z}\bar{z}}\frac{\partial
h_{\bar{z}\bar{z}}}{\partial\bar{z}}+\mathcal{O}(\epsilon^2),\nno\\
\Ga^z_{zz}&=&4(1-4\epsilon-2\epsilon\ln(z\bar{z}))h_{z{\bar{z}}}\frac{\partial
h_{z{\bar{z}}}}{\partial z}+\frac{1}{2}h^{zz}\frac{\partial
h_{zz}}{\partial z}+\mathcal{O}(\epsilon^2),\nno\\
\Ga^{\bar{z}}_{\rho\bar{z}}&=&\Ga^z_{\rho
z}=\frac{1}{l_a}(1-4\epsilon-2\epsilon\ln(z\bar{z}))+\mathcal{O}(\epsilon^2)
,\nno\\
\Ga^\rho_{z\bar{z}}&=&\Ga^\rho_{\bar{z}z}=-e^{2\rho/l_a}\frac{h_{z\bar{z}}}{l_a}\quad
{\rm and}\quad
\Ga^z_{\bar{z}\rho}=-\frac{4}{l_a}e^{-2\rho/l_a}(1-4\epsilon-2\epsilon\ln(z\bar{z}))h_{\bar{z}\bar{z}}h_{z\bar{z}}+\mathcal{O}(\epsilon^2),
 \nno\\ \Ga^{\bar{z}}_{\rho
 z}&=&-\frac{4}{l_a}e^{-2\rho/l_a}(1-4\epsilon-2\epsilon\ln(z\bar{z}))h_{zz}h_{\bar{z}z}+\mathcal{O}(\epsilon^2),
 \nno\\ \Ga^{\bar{z}}_{zz}&=&-4e^{-2\rho/l_a}(1-4\epsilon-2\epsilon\ln(z\bar{z}))h_{zz}\frac{\partial h_{z{\bar{z}}}}{\partial
z}+\frac{1}{2}g^{\bar{z}z}\frac{\partial h_{zz}}{\partial
z}+\mathcal{O}(\epsilon^2),\nno\\
\Ga^{z}_{\bar{z}\bar{z}}&=&-4e^{-2\rho/l_a}(1-4\epsilon-2\epsilon\ln(z\bar{z}))h_{\bar{z}\bar{z}}\frac{\partial
h_{z{\bar{z}}}}{\partial
\bar{z}}+\frac{1}{2}g^{z\bar{z}}\frac{\partial
h_{\bar{z}\bar{z}}}{\partial \bar{z}}+\mathcal{O}(\epsilon^2).\ee
From Eq.(\ref{sr}), the singular part of the Riemann tensor and
Riemann scalar are
\be\label{sr1} ^sR^z_{zz\bar{z}}
&=&^sR^{\bar{z}}_{\bar{z}\bar{z}z}=-2\pi\epsilon\delta^{(2)}(z,\bar{z})\quad
{\rm and}\quad ^sR^{\bar{z}}_{zz\bar{z}}=
^sR^z_{\bar{z}\bar{z}z}=4\pi\epsilon
e^{-2\rho/l_a}\delta^{(2)}(z,\bar{z}),
 \nno\\ ^sR&=&-8\pi\epsilon
e^{-2\rho/l_a}\delta^{(2)}(z,\bar{z})+32\pi\epsilon
e^{-6\rho/l_a}\delta^{(2)}(z,\bar{z})+\mathcal{O}(\epsilon^2).\ee
Like Eq.(\ref{EH1}), the Einstein-Hilbert action is divided into two
parts
\be I_{EH}&=& ^rI_{EH}+ \frac{1}{16\pi
G}\int_{\mathcal{M}}\sqrt{-g}dz d\bar{z}dx
 ^sR, \ee
and the contribution to the HEE is
\be
S_{EH}\label{seh}&=&\frac{1}{4G}\int_0^{\rho_0}d\rho-\frac{1}{G}\int_0^{\rho_0}e^{-4\rho/l_a}d\rho
 \nno\\
&=&\frac{c}{6}\ln\frac{L}{a}-\frac{c}{6}+\mathcal{O}((\frac{a}{L})^4)\simeq\frac{c}{6}\ln\frac{L}{a}-\frac{c}{6}.
\ee
The first term is just the universal term of the HEE obtained in
\cite{cardy,rt,fursaev}~\footnote{Compared to the calculation in
field theory side, there is a difference by a factor $2$. This is
due to the fact that here we only considered one brunch point, while
there are two brunch points at the boundary between $A$ and $B$.
With this  Eq.(\ref{seh}) can reproduce the correct result.}. Here
$\rho_0$ is related to the UV cutoff $a$ of the dual 2-dimensional
CFT by $\exp(\rho_0/l_a)=L/a$, and $L$ is the length of the whole
system $A+B$.

To compute the correction of the gravitational Chern-Simons term to
HEE, we can use the result in \cite{offshell,solo}, the introduction
of the singular coordinate transformation in eq.(\ref{w}) causes
conical singularities on the bulk minimal surface. As a result, the
induced Riemann tensor on the minimal surface is
\be\label{Riemann} R^{\mu\nu}_{~~\al\beta}&=&
^rR^{\mu\nu}_{~~\al\beta}+2\pi\epsilon[(\hat n^{\mu}\hat n_{\al})
(\hat n^{\nu}\hat n_{\beta})-(\hat n^{\mu}\hat n_{\beta}) (\hat
n^{\nu}\hat n_{\al})]\delta_\Si.\ee
where $\delta_\Si$ is a 2-d delta function which is nonzero on the
bulk minimal surface. $\hat n_1$ and $\hat n_2$ are the unit normal
vectors of $\Si$, and $\hat n^{\mu}\hat n_{\al}=\hat n_1^{\mu}\hat
n^1_{\al}+\hat n_2^{\mu}\hat n^2_{\al}$. Then the Chern-Simons term
is also divided into two parts
\be\label{CSsingular1}I_{CS}&=&^rI_{CS}-\frac{\al\epsilon}{32
G}\int_{\mathcal{M}}d^3x\sqrt{-g}\Ga_{\al\beta\mu}\epsilon^{\mu\nu\la}
[(\hat n^{\beta}\hat n_{\nu}) (\hat n^{\al}\hat n_{\la})-(\hat
n^{\beta}\hat n_{\la}) (\hat n^{\al}\hat n_{\nu})]\delta_\Si,\ee
where $^rI_{CS}$ is the regular part of the gravitational
Chern-Simons term. From the Appendix, the singular part of
eq.(\ref{CSsingular1}) is
\be\label{cssingular4}^sI_{CS}=\frac{\al\epsilon}{32 G}\int_\Si
dx\sqrt{\ga}\frac{\epsilon^{z{\bar
z}\rho}}{2}(h_{tt}-h_{xx})_{,\rho}({\hat n}
^{\rho}_1)^2\frac{h_{tx}}{h_{tt}}\frac{1}{|\frac{dF}{d\rho}|}.\ee
Since $h_{tx}$ is zero in eq.(\ref{gads}), $^sI_{CS}$ vanishes,
consequently, $S_{CS}=0$. Thus we can see that the gravitational
Chern-Simons term does not correct the HEE of the 2-d CFT (at zero
temperature) on the boundary of global AdS$_3$ spacetime.

\subsection{Finite temperature CFT case}

 The finite temperature 2-dimensional CFT case corresponds to the case of the bulk
 having a BTZ black hole with metric~\cite{btz}
\be\label{btz}
ds^2=-\frac{(r^2-r_+^2)(r^2-r_-^2)}{l_a^2r^2}dt^2+\frac{l_a^2r^2}{(r^2-r_+^2)(r^2-r_-^2)}dr^2
+r^2(d\phi-\frac{r_+r_-}{l_ar^2}dt)^2. \ee
The black hole is of mass $m=(r_+^2+r_-^2)/(8Gl_a^2)$, angular
momentum $j=r_+r_-/(4Gl_a)$ and temperature $T=(r_+^2-r_-^2)/(2\pi
r_+l_a^2)$. It can also be expressed in the Gaussian normal
coordinates by setting
\be d\rho=\frac{l_ardr}{\sqrt{(r^2-r_+^2)(r^2-r_-^2)}},\ee
which gives
\be l_ae^{\rho/l_a}=\sqrt{r^2-r_+^2}+\sqrt{r^2-r_-^2}.\ee
In the $\rho\gg l_a$ limit, we have
\be
e^{2\rho/l_a}&=&\frac{4r^2}{l_a^2}-16Gm-\frac{16G^2l_a^2m^2}{r^2}+\frac{16G^2j}{r^2}
+\mathcal{O}(r^{-4}) \nno\\ \quad {\rm and}\quad
r^2&=&\frac{l_a^2}{4}e^{2\rho/l_a}+4l_a^2Gm+4e^{-2\rho/l_a}(4G^2l_a^2m^2-G^2j^2).\ee
Then the metric can be expanded as
\be \label{btzfg} ds^2&=&d\rho^2+(-\frac{1}{4}e^{2\rho/l_a}+4Gm)dt^2
+(\frac{l_a^2}{4}e^{2\rho/l_a}+4Gl_a^2m+4e^{-2\rho/l_a}(4G^2l_a^2m^2-G^2j^2))d\phi^2
 \nno \\ &&-8Gjd\phi dt-4e^{-2\rho/l_a}(4G^2m^2-\frac{G^2j^2}{l_a^2})dt^2+\mathcal{O}(e^{-4\rho/l_a})\nno\\
&=&d\rho^2+e^{2\rho/l_a}(-\frac{1}{4}dt^2+\frac{l_a^2}{4}d\phi^2)+4Gmdt^2+4Gl_a^2md\phi^2
-8Gjd\phi dt \nno\\
&&+e^{-2\rho/l_a}((4\frac{G^2j^2}{l_a^2}-16G^2m^2)dt^2+16G^2l_a^2m^2d\phi^2-4G^2j^2d\phi^2)
+\mathcal{O}(e^{-4\rho/l_a}).\ee
In this case, we need to take $t\rightarrow -i\tau$, $j\rightarrow
ij$ and define
\be w=\frac{1}{2}\tau+i\frac{l_a}{2}\phi=z^n=z^{1+\epsilon}.\ee
After reserving the $h^{(2)}_{ij}$ terms we obtain
\be\label{btzfg1}
ds^2&=&d\rho^2+2e^{2\rho/l_a}h_{z\bar{z}}dzd\bar{z}
+(i\frac{8Gj}{l_a}-8Gm)h_{zz}dz^2-(i\frac{8Gj}{l_a}+8Gm)h_{\bar{z}\bar{z}}d\bar{z}^2
+\mathcal{O}(\epsilon^2),\ee
where $h_{z\bar{z}}$, $h_{zz}$ and $h_{\bar{z}\bar{z}}$ are the same
as those in the pure AdS$_3$ case. Thus just following the procedure
above, the singular part of the Riemann tensors and Ricci scalar are
\be\label{sr2} ^sR^z_{zz\bar{z}}
&=&^sR^{\bar{z}}_{\bar{z}\bar{z}z}=-2\pi\epsilon\delta^{(2)}(z,\bar{z}),
\nno\\ ^sR^{\bar{z}}_{zz\bar{z}}&=& 4\pi\epsilon
e^{-2\rho/l_a}(i\frac{8Gj}{l_a}-8Gm)\delta^{(2)}(z,\bar{z})\quad
{\rm and}\quad ^sR^z_{\bar{z}\bar{z}z}=-4\pi\epsilon
e^{-2\rho/l_a}(i\frac{8Gj}{l_a}+8Gm)\delta^{(2)}(z,\bar{z}),
 \nno\\ ^sR&=&-8\pi\epsilon
e^{-2\rho/l_a}\delta^{(2)}(z,\bar{z})+32\pi\epsilon
e^{-6\rho/l_a}((8Gm)^2+(\frac{8Gj}{l_a})^2)\delta^{(2)}(z,\bar{z})+\mathcal{O}(\epsilon^2).
\ee
Then the total action is
\be\label{abtz} I&\simeq& ^rI_{EH}-\frac{\epsilon}{4
G}\int_{\rho_+}^{\rho_0} d\rho+\frac{\epsilon}{
G}((8Gm)^2+(\frac{8Gj}{l_a})^2)\int_{\rho_+}^{\rho_0}
e^{-4\rho/l_a}d\rho + ^rI_{CS}\nonumber\\&&+\frac{\al\epsilon
j}{4l_a^2 }\int_\Si dx\sqrt{\ga}\frac{(\frac{\partial F}{\partial
\rho})^2}{g^{xx}(\frac{\partial F}{\partial x})^2+(\frac{\partial
F}{\partial \rho})^2}\frac{e^{-2\rho/l_a}}{|\frac{dF}{d\rho}|}\nno\\
 &=& ^rI_{EH}+ ^rI_{CS}-\frac{c\epsilon}{6}\ln\frac{\beta
}{a}+\frac{c\epsilon}{6}\ln\frac{\sqrt{r_+^2-r_-^2}}{l_a}
-\frac{c\epsilon}{6}((8Gm)^2+(\frac{8Gj}{l_a})^2)(\frac{a^4}{\beta^4}-\frac{1}{4\pi^2
r_+^2T^2})\nonumber\\&&+\frac{\al\epsilon j}{4l_a^2 }\int_\Si
dx\sqrt{\ga}\frac{(\frac{\partial F}{\partial
\rho})^2}{g^{xx}(\frac{\partial F}{\partial x})^2+(\frac{\partial
F}{\partial \rho})^2}\frac{e^{-2\rho/l_a}}{|\frac{dF}{d\rho}|} \nonumber \\
&\simeq& ^rI_{EH}+
^rI_{CS}-\frac{c\epsilon}{6}\ln\frac{\beta}{a}+\frac{c\epsilon}{12}\ln
(2\pi
r_+T)+\frac{c\epsilon}{6}((8Gm)^2+(\frac{8Gj}{l_a})^2)\frac{1}{4\pi^2
r_+^2T^2}\nno\\&&+\frac{\al\epsilon j}{4l_a^2 }\int_\Si
dx\sqrt{\ga}\frac{(\frac{\partial F}{\partial
\rho})^2}{g^{xx}(\frac{\partial F}{\partial x})^2+(\frac{\partial
F}{\partial \rho})^2}\frac{e^{-2\rho/l_a}}{|\frac{dF}{d\rho}|}.\ee
Here we have taken the UV cutoff $e^{\rho_0/l_a} \sim \beta /a$ as
in \cite{rt} and $\beta $ is the inverse of the temperature $T$.
Thus we can obtain the HEE of the dual 2-dimensional CFT with
gravitational Chern-Simons correction at finite temperature $T$
\be
\label{hee2}S_A&\simeq&\frac{c}{3}\ln\frac{\beta}{a}-\frac{c}{6}\ln
(2\pi r_+ T)-\frac{c}{3}((8Gm)^2+(\frac{8Gj}{l_a})^2)\frac{1}{4\pi^2
r_+^2T^2}\nno\\ &&-\frac{\al r_+ r_-}{16Gl_a^3 }\int_\Si
dx\frac{\frac{\partial F}{\partial \rho}\sqrt{(\frac{\partial
F}{\partial x})^2+g_{xx}(\frac{\partial F}{\partial
\rho})^2}}{g^{xx}(\frac{\partial F}{\partial x})^2+(\frac{\partial
F}{\partial \rho})^2}\frac{e^{-2\rho/l_a}}{|\frac{dF}{d\rho}|}.\ee
Generally speaking, we need to solve the equation for the bulk
static minimal surface to get the explicit contribution of the last
term in eq.(\ref{hee2}). For simplicity, here we would like to
evaluate this term by making some approximations. Since the CFT
system is located at the circle with radius $\rho=\rho_0\gg l_a$,
the bulk static minimal surface now is just a minimal curve which
connect the boundary (two points) between subsystems $A$ and $B$.
The length of the minimal curve should have the same order of the
arc length $l_A$ of the subsystem $A$. When $l_A$ is small compared
to the length of the boundary, the bulk minimal curve is close to
the arc of $A$ on the boundary. Thus, the equation for the bulk
minimal curve can be approximately evaluated by
\be\label{eqmini}F=\rho(x)-\rho_0.\ee
So $dF/d\rho=\partial F/\partial\rho=1$ and $\partial F/\partial
x=0$. Consequently, we obtain the gravitational Chern-Simons term
correction to HEE
\be \label{hee3}S_{CS}&\sim&-\frac{\al \pi r_+
r_-}{16Gl_a^2}\frac{a}{\beta}.\ee
It is obvious that $S_{CS}$ is an finite correction to the HEE of
subsystem $A$. Now let us consider two special cases. One is the
extremal black hole case or zero temperature limit with $r_+=r_-$,
the other is the nonrotating BTZ black hole case with
$r_-\rightarrow 0$. \\
When $r_+=r_-$, $\beta\rightarrow \infty$, we have
\be
\label{hee3}S_A&\simeq&\frac{c}{3}\ln\frac{L}{a}-\frac{8c}{3}\frac{r_+^4}{l_a^4}.\ee
Note that here we have restored the UV cutoff $e^{\rho_0/l_a} \sim
L/a$ and $\rho\in[0, \rho_0]$ in the zero temperature limit. We can
see that in this zero temperature limit, the gravitational
Chern-Simons term correction to HEE vanishes and the universal part of
HEE is the same as that in pure AdS$_3$ spacetime.\\
When $r_-\rightarrow 0$, we have
\be
\label{hee4}S_A&\simeq&\frac{c}{3}\ln\frac{\beta}{a}-\frac{c}{3}\ln\frac{r_+}{l_a}-\frac{c}{3}.
\ee
The HEE computed directly from the bulk minimal surface in
nonrotating BTZ black hole metric is \cite{rt}
\be\label{heenbtz} S_A \simeq\frac{c}{3}\ln(\frac{\beta}{\pi a}\sinh
\frac{\pi l_A}{\beta})=\frac{c}{3}\ln(\frac{\beta }{\pi
a})+\frac{c}{3}\ln(\sinh \frac{r_+ l_A}{2l_a^2}).\ee
Comparing eq.(\ref{hee4}) with eq.(\ref{heenbtz}), the leading terms
give the same result, up to a small finite term $\sim \ln \pi$. In
the low temperature phase, the second term in eq.(\ref{heenbtz})
gives $\frac{c}{3}\ln \frac{r_+}{l_a}$, while the second term in
eq.(\ref{hee4}) gives a negative contribution $-\frac{c}{3}\ln
\frac{r_+}{l_a}$, although both terms are small compared to the
leading term. Also, like the zero temperature case, the contribution
from the gravitational Chern-Simons term to HEE is zero up to the
order of $h^{(2)}_{ij}$. From eq.(\ref{sbtz}), we can see that when
$r_-\rightarrow 0$, the nonrotating BTZ black hole entropy is also
not corrected due the bulk gravitational Chern-Simons term.

\section{Conclusions and Discussions}

In this paper, we studied the gravitational Chern-Simons term
correction to the HEE of the dual 2-dimensional CFT on the conformal
boundary of AAdS$_3$ spacetime by using the off-shell Euclidean path
integral approach. As we have expected, like the BTZ black hole
entropy, although the bulk geometry does not change in the presence
of the gravitational Chern-Simons term, the HEE is indeed modified
by an finite term (up to the order of $h^{(2)}_{ij}$) at finite
temperature case  where the bulk is the BTZ black hole. This is
because the presence of the gravitational Chern-Simons term violates
the bulk diffeomorphism invariance of the gravitational action. In
the AdS/CFT correspondence, the violation causes the gravitational
anomaly (covariant anomaly) in the dual 2-dimensional CFT on the
conformal boundary. Since the covariant anomaly modifies the central
charges of CFT, it also should make a contribution to the HEE. While
in the zero temperature and nonrotating BTZ black hole cases, the
gravitational Chern-Simons term does not contribute to HEE in our
calculation. However, the central charges of the dual CFT are
corrected, too. A possible explanation is that the large $N$
correction caused by the bulk gravitational Chern-Simons term is not
high enough to excite the NS-NS vacuum (since there is a mass gap
between AdS$_3$ spacetime and the black hole spectrum) and the dual
field of nonrotating BTZ black hole, so its entropy does not change.
A support of this point is that from eq.(\ref{sbtz}), when $r_-=0$,
the nonrotating BTZ black hole entropy is not corrected. Anyway,
this point needs a further study. It can be seen that, when the
temperature $T$ is zero or finite, our results in the case without
gravitational Chern-Simons term correction reproduce those obtained
from the geometric calculation. It should be stressed here that in
the calculation, we did not actually solve the constraint equation
$\delta \mathcal{A}=0$. Instead, we just used the approximate UV-IR
relations $\exp(\rho_0/l_a)=L/a$ at zero temperature and
$\exp(\rho_0/l_a)=\beta/a$ at finite temperature, these relations
are enough to show the gravitational Chern-Simons term correction to
HEE in the leading order. To find the full exact result, one need to
solve $\delta \mathcal{A}=0$ and find out the exact UV-IR relation.

We want to point out that generally, the HEE is not the same as the
black hole entropy when there is a black hole in the bulk. In fact,
as has been shown in \cite{covhee}, the bulk minimal surfaces are
generated either by null curves (with black hole in bulk) or
spacelike curves (without black hole in bulk), and the area of the
bulk minimal surface acts as the entropy bound (such as the
covariant entropy bound \cite{bousso}) of the CFT system. Thus the
HEE indicates the maximum entropy of the CFT system.


\begin{acknowledgments}\vskip -4mm

JRS would like to thank R.-G. Cai, D. Grumiller, C.-G. Huang, T.
Takayanagi and Y. Tian for very valuable discussions and comments.
Useful discussions with L.-M. Cao, Z. Chang, and D.-W. Pang are also
acknowledged. This work is partly supported by NSFC (Grant Nos
90503002, 10775140 and 10701081), Knowledge Innovation Funds of CAS
(KJCX3-SYW-S03), NKBRPC(2004CB318000), and Beijing Jiao-Wei Key
project(KZ200810028013).

\end{acknowledgments}

\section{Appendix}
\begin{appendix}

For AdS$_3$ spacetime and BTZ black hole, eq.(\ref{Gauss}) can be
expressed as
\be\label{Gaussnormal}ds^2=d\rho^2+g_{tt}dt^2+2g_{tx}dtdx+g_{xx}dx^2,\ee
where $g_{tt}=h_{tt}$, $g_{tx}=h_{tx}$, $g_{xx}=h_{xx}$ and
$g=h\simeq -\frac{1}{4}e^{4\rho/l_a}$. Set the bulk static minimal
surface (denoted by $\Si$) be
\be\label{ms}x=x(\rho)\quad {\rm or}\quad F[x(\rho),\rho]=0. \ee
One of its normal vectors is
\be\label{nv1}n^1=(0, \frac{\partial F}{\partial x}, \frac{\partial
F}{\partial \rho}).\ee
There is another normal vector of the bulk static minimal surface
\be\label{nv2}n^2=(f,\tilde{f},0),\ee
where $f$ and $\tilde{f}$ are some function which can be determined
by the normalization condition. From eq.(\ref{nv1}) and
eq.(\ref{nv2}) we can read the nonvanishing components of these
normal vectors
\be\label{nvc}n^1_x=\frac{\partial F}{\partial x},\quad
n^1_\rho=\frac{\partial F}{\partial \rho},\quad n^2_t=f,\quad
n^2_x=\tilde{f}.\ee
Thus we have
\be\label{n1n1}n^1n_1&=&n^1_t n^t_1+n^1_x n^x_1+n^1_\rho
n^\rho_1=n^1_x(g^{xx}n_{1x}+g^{xt}n_{1t})+g^{\rho\rho}n^\rho_1
n^{1\rho}\nno\\&=&g^{xx}(\frac{\partial F}{\partial
x})^2+(\frac{\partial F}{\partial \rho})^2.\ee
and
\be\label{n2n2}n^2n_2=n^2_t n^t_2+n^2_x
n^x_2=f^2g^{tt}+2f\tilde{f}g^{tx}+\tilde{f}^2g^{xx}.\ee
Using the normalization conditions $\hat{n}^1\hat{n}_1=1$,
$\hat{n}^2\hat{n}_2=-1$ and $\hat{n}^2\hat{n}_1=0$ we get the unit
normal vectors
\be\label{nvunit}\hat{n}^1&=&\frac{1}{\sqrt{g^{xx}(\frac{\partial
F}{\partial x})^2+(\frac{\partial F}{\partial \rho})^2}}(0,
\frac{\partial F}{\partial x}, \frac{\partial F}{\partial
\rho}),\quad {\rm and}\quad
\hat{n}^2=(\sqrt{-g_{tt}},\frac{-g_{tx}}{\sqrt{-g_{tt}}},0),\nno\\
\hat{n}_1&=&\frac{1}{\sqrt{g^{xx}(\frac{\partial F}{\partial
x})^2+(\frac{\partial F}{\partial \rho})^2}}(g^{tx}\frac{\partial
F}{\partial x}, g^{xx}\frac{\partial F}{\partial x}, \frac{\partial
F}{\partial \rho}),\quad {\rm and}\quad
\hat{n}_2=\frac{1}{\sqrt{-g_{tt}}}(-1, 0, 0),\ee

The singular part of the gravitational Chern-Simons term is
\cite{solo}
\be\label{CSsingular2}^sI_{CS}&=&-\frac{\al\epsilon}{32
G}\int_{\mathcal{M}}d^3x\sqrt{-g}\Ga_{\al\beta\mu}\epsilon^{\mu\nu\la}
[(\hat n^{\beta}\hat n_{\nu}) (\hat n^{\al}\hat n_{\la})-(\hat
n^{\beta}\hat n_{\la}) (\hat n^{\al}\hat n_{\nu})]\delta_\Si,\ee
where
\be\label{deltaSi}
\delta_\Si=\frac{\delta(t-C)\delta(F[x(\rho),\rho])}{{\rm
measure\quad of\quad the}\quad t, \rho\quad{\rm components}}\ee
is the 2-d delta function which is nonvanishing only on the bulk
static minimal surface $\Si$, $C$ is some const. time and
\be\label{deltaF}\delta(F[x(\rho),\rho])=\frac{\delta(\rho-\bar\rho)}{|\frac{dF}{d\rho}|},\ee
where $\bar\rho$ is the value of $\rho$ on $\Si$. Integrating out
the delta function in eq.(\ref{CSsingular2}), we obtain
\be\label{cssingular3}^sI_{CS}&=&-\frac{\al\epsilon}{32 G}\int_\Si
dx\sqrt{\ga}\Ga_{\al\beta\mu}\epsilon^{\mu\nu\la} [(\hat
n^{\beta}\hat n_{\nu}) (\hat n^{\al}\hat n_{\la})-(\hat
n^{\beta}\hat n_{\la}) (\hat n^{\al}\hat
n_{\nu})]\frac{1}{|\frac{dF}{d\rho}|},\ee
where $\sqrt{\ga}dx=\sqrt{(d\rho/dx)^2+g_{xx}}dx$ is the invariant
measure on $\Si$.

Making coordinate transformation in the Gaussian normal coordinate
eq.(\ref{Gaussnormal}), labeling $x^0=t=-i\tau$, $j\rightarrow ij$
and $x^1=x$. Defining
\be\label{cfmflat} w=\frac{1}{2}(\tau+ix) \quad {\rm and} \quad
\bar{w}=\frac{1}{2}(\tau-ix),\ee
then we have $\tau=w+\bar{w}$ and $x=-i(w-\bar{w})$. Thus we can
rewrite eq.(\ref{Gaussnormal}) as
\be\label{cf}
ds^2=d\rho^2-(h_{00}+2h_{01}+h_{11})dw^2-2(h_{00}-h_{11})dwd\bar{w}
-(h_{00}-2h_{01}+h_{11})d\bar{w}^2. \ee
Like before, we further introduce a singular coordinate
transformation $w=z^n=z^{1+\epsilon}$ and expand it in the linear
order of $\epsilon$. After that we obtain
\be\label{cf1}ds^2&=&d\rho^2-(h_{00}+2h_{01}+h_{11})h_{zz}dz^2
-4(h_{00}-h_{11})h_{z\bar{z}}dzd\bar{z}
-(h_{00}-2h_{01}+h_{11})h_{\bar{z}\bar{z}}d\bar{z}^2\nno\\
& \equiv &
d\rho^2+g_{zz}(\rho,z)dz^2+2g_{z\bar{z}}(\rho,z,\bar{z})dzd\bar{z}
+g_{\bar{z}\bar{z}}(\rho,\bar{z})d\bar{z}^2. \ee
Then the unit normal vectors $\hat{n}_1$ and $\hat{n}_2$ become
\be\label{nvunitE} \hat{n}_1&=&\frac{1}{\sqrt{g^{xx}(\frac{\partial
F}{\partial x})^2+(\frac{\partial F}{\partial
\rho})^2}}[\frac{i}{2}(g^{tx}\frac{\partial F}{\partial
x}+g^{xx}\frac{\partial F}{\partial
x})\partial_z+\frac{i}{2}(g^{tx}\frac{\partial F}{\partial
x}-g^{xx}\frac{\partial F}{\partial x})\partial_{\bar
z}+\frac{\partial F}{\partial \rho}\partial_\rho],\nno\\
\hat{n}_2&=&\frac{-i}{\sqrt{-g_{tt}}}[\frac{1}{2}\partial_z+\frac{1}{2}
\partial_{\bar z}],\ee
The nonvanishing components of $\Ga_{\al\beta\la}$ are
\be\label{christoffel}\Ga_{z{\bar
z}z}&=&-\frac{1}{2}g_{\bar{z}z,z},\quad \Ga_{z{\bar z}{\bar
z}}=\frac{1}{2}g_{\bar{z}z,\bar z},\quad \Ga_{z\rho
z}=\frac{1}{2}g_{zz,\rho},\nno\\ \Ga_{z\rho\bar
z}&=&\frac{1}{2}g_{z{\bar z},\rho},\quad \Ga_{{\bar z}\rho
z}=\frac{1}{2}g_{z{\bar z},\rho},\quad \Ga_{{\bar z}\rho\bar
z}=\frac{1}{2}g_{{\bar z}{\bar z},\rho}.\ee
For AdS$_3$ spacetime and BTZ black hole, up to the order of
$h^{(2)}$, $g_{zz,\rho}=0$ and $g_{{\bar z}{\bar z},\rho}=0$.
Meanwhile, both $g_{\bar{z}z,z}$ and $g_{\bar{z}z,\bar z}$ contain
an $\epsilon$ factor, thus they do not contribute to $^sI_{CS}$ up
to the order of $\epsilon$. Substituting eq.(\ref{nvunitE}) and
eq.(\ref{christoffel}) into eq.(\ref{cssingular3}), we obtain
\be\label{cssingular4}^sI_{CS}=\frac{\al\epsilon}{32 G}\int_\Si
dx\sqrt{\ga}\frac{\epsilon^{z{\bar
z}\rho}}{2}(h_{tt}-h_{xx})_{,\rho}({\hat n}
^{\rho}_1)^2\frac{h_{tx}}{h_{tt}}\frac{1}{|\frac{dF}{d\rho}|}.\ee
For the AdS$_3$ spacetime, eq.(\ref{cssingular4}) vanishes since
$h_{tx}$ is zero, i.e., the gravitational Chern-Simons term does not
alter the HEE in the zero temperature case. While for the BTZ black
hole, up to the order of $h^{(2)}$, we have
\be\label{cssingularbtz}^sI_{CS}=\frac{\al\epsilon j}{4l_a^2
}\int_\Si dx\sqrt{\ga}\frac{(\frac{\partial F}{\partial
\rho})^2}{g^{xx}(\frac{\partial F}{\partial x})^2+(\frac{\partial
F}{\partial \rho})^2}\frac{e^{-2\rho/l_a}}{|\frac{dF}{d\rho}|}.\ee
With the help of eq.(\ref{sa3}), we obtain the gravitational
Chern-Simons term correction to HEE at the finite temperature case
\be\label{cscorrectbtz}S_{CS}=-\frac{\al r_+ r_-}{16Gl_a^3 }\int_\Si
dx\frac{\frac{\partial F}{\partial \rho}\sqrt{(\frac{\partial
F}{\partial x})^2+g_{xx}(\frac{\partial F}{\partial
\rho})^2}}{g^{xx}(\frac{\partial F}{\partial x})^2+(\frac{\partial
F}{\partial \rho})^2}\frac{e^{-2\rho/l_a}}{|\frac{dF}{d\rho}|}.\ee
In the above calculation, we have used the following result: \\
Under general coordinate transformation $x\rightarrow x'$,
\be \sqrt{-g}\epsilon^{\mu\nu\la}d^3x&\rightarrow&
\sqrt{-g'}\epsilon'^{\mu\nu\la}d^3x'\nno\\
&&=|\frac{\partial x}{\partial x'}|\sqrt{-g}\frac{\partial
x'^{\mu}}{\partial x^{\al}}\frac{\partial x'^{\nu}}{\partial
x^{\beta}}\frac{\partial x'^{\la}}{\partial
x^{\ga}}\epsilon^{\al\beta\ga}|\frac{\partial x'}{\partial x}|d^3x,
\ee
where $g'$=$4g$ in this case, which gives
\be \epsilon^{\rho z\bar{z}}&=&\frac{i}{2\sqrt{-g}}, \ee
and
\be \sqrt{-g'}\epsilon^{\rho z\bar{z}}d\rho
dzd\bar{z}=2\sqrt{-g}\frac{i}{2\sqrt{-g}}d\rho dzd\bar{z}=id\rho
dzd\bar{z}.\ee
\end{appendix}

\end{document}